\documentclass[fleqn,3p,number,11pt]{article}

\usepackage[a4paper, total={18cm, 26cm}]{geometry}

\usepackage[utf8]{inputenc}

\usepackage{amsmath, amsthm, amsfonts, amssymb}
\usepackage{bm}
\usepackage{todonotes}
\usepackage{xcolor}
\usepackage{subcaption}

\newcommand{\displacement}{{\bm u}}
\newcommand{\energy}{\mathcal{E}}
\newcommand{\pf}{\varphi}
\newcommand{\D}{\mathcal{D}}

\newcommand{\R}{\mathcal{R}}

\DeclareMathOperator*{\argmin}{arg\,min}

\usepackage{tikz}
\usepackage{pgfplots,pgfplotstable}
\pgfplotsset{compat=newest}
\definecolor{blau0} {RGB}{ 131 198 216} 
\definecolor{grau}  {RGB}{  0  84 159}
\definecolor{rot}   {RGB}{204   7  30}
\definecolor{blau2} {RGB}{  0  61 128}
\definecolor{grun2} {RGB}{  0 85   0}
\definecolor{rot2}  {RGB}{120   7  30}
\definecolor{gelb}  {RGB}{70 70 70}

\definecolor{blau}{RGB}{0 144 188}
\definecolor{newblue1}{RGB}{0 144 188}
\definecolor{newblue2}{RGB}{197 216 227}
\definecolor{newgreen1}{RGB}{0 144 118}
\definecolor{grun}{RGB}{255 137 0}
\definecolor{newgreen2}{RGB}{197 222 215}
\definecolor{neworange1}{RGB}{255 137 0}
\definecolor{neworange2}{RGB}{255 205 105}

\usepackage{authblk}

\title{A Cahn-Hilliard-Biot system and its generalized gradient flow structure}

\author[1]{Erlend Storvik\footnote{Corresponding author: erlend.storvik@uib.no}}
\author[1]{Jakub Wiktor Both}
\author[1]{Jan Martin Nordbotten}
\author[1]{Florin Adrian Radu}

\affil[1]{Department of Mathematics, University of Bergen, All\'egaten 44, 5007 Bergen, Norway}

\date{}

\begin{document}

\maketitle

\begin{abstract}
In this work, we propose a new model for flow through deformable porous media, where the solid material has two phases with distinct material properties. The two phases of the porous material follow a Cahn-Hilliard type evolution, with additional impact from both elastic and fluid effects, and the coupling between flow and deformation is governed by Biot's theory. This results in a three-way coupled system which can be seen as an extension of the Cahn-Larch\'e equations with the inclusion of a fluid flowing through the medium. The model covers essential coupling terms for several relevant applications, including solid tumor growth, biogrout, and wood growth simulation. Moreover, we show that this coupled set of equations follow a generalized gradient flow framework. This opens a toolbox of analysis and solvers which can be used for further study of the model. Additionally, we provide a numerical example showing the impact of the flow on the solid phase evolution in comparison to the Cahn-Larch\'e system.
\end{abstract}

\section{Introduction\label{sec:intro}}
In this letter, we develop a general model with the ability to capture situations with flow through a deformable porous medium that changes character in terms of stiffness, permeability, compressibility, and poroelastic coupling strength due to Cahn-Hilliard-type phase changes in the solid matrix. There are several applications where this type of behavior exists. One example being solid tumor evolution, where it is argued that stress effects resulting from tumor growth impact the tumor evolution itself  
\cite{lima2016, lima2017},
and that stress can inhibit tumor growth \cite{tumorstresscheng,tumorstresshelminger, tumorstressstylianopoulos}. Moreover, the elastic properties of the surrounding. matrix and the interstitial fluid pressure are elevated in most solid malignant tumors \cite{milosevictumor}. One can then consider the two-phase porous medium as cancerous and healthy cells with the surrounding extracellular matrix, and the fluid as the interstitial fluid. Similar models involving Cahn-Hilliard-type evolution of tumor growth can be seen in \cite{fritz2021analysis,fritz2020subdiffusive, garcketumormechanics, ebenbeck2019analysis, cristini2010book, oden2010mixture, cristini2009nonlinear, colli2015vanishCH, fritz2019localnonlocal, garcke2018darcy, krejci2021deformtumorflow}. Additional applications of poroelastic media with solid phase changes range from biogrout to wood growth, where sapwood transforms to heartwood.

The proposed system is an extension of the Cahn-Hilliard model and the quasi-static linear Biot equations, where the Cahn-Hilliard contribution governs the solid phase changes in the system through a smooth phase-field variable, and the Biot equations govern flow and elasticity. The Cahn-Hilliard equation originates from the work of Cahn and Hilliard \cite{CH1}, where the interfacial free energy of a non-uniform composition was introduced to model phase separation. Coupling the Cahn-Hilliard model with elasticity, is often called the Cahn-Larch\'e model due to its origination \cite{CL}, and several applications have been considered with this model in mind, including li-ion batteries \cite{areias}, and tumor evolution \cite{garcketumormechanics,fritz2020subdiffusive}. In this work, we include fluid in the system, which is assumed to flow through the poroelastic medium with Biot-type coupling between flow and elasticity \cite{coussy}. 

We show that the resulting model has a generalized gradient flow structure, i.e., a dissipative system where the state of the system evolves with the negative gradient of its free energy. The extension to generalized gradient flows allows for non-quadratic, and partially degenerate, dissipation potentials, and there is currently an increasing interest in the mathematics of generalized gradient flows, both with respect to modeling \cite{peletierlecturenotes, cances2015gradient}, abstract analysis \cite{colli1992, savare2000, bartels2014, jakubgradientflow} and numerical solution strategies \cite{jakubgradientflow, jungel2019}. It is long known that the Cahn-Hilliard equation and single-phase flow through porous media can be written as standard gradient flows, and it was showed in \cite{jakubgradientflow} that the Biot equations have a generalized gradient flow structure. Here, we show that even though it is not obvious that the combination of two gradient flows retains the structure, the Cahn-Hilliard-Biot model does, indicating the thermodynamical consistency of the model. This will be a valuable toolbox for further study and development of mathematics for the model, both with respect to well-posedness analysis and numerical solution strategies.


The letter is structured as follows: In Section~\ref{sec:model}, the Cahn-Hilliard-Biot model is presented. Conservation laws for each of the three coupled processes; phase-field evolution, elasticity, and fluid flow are introduced, then the free energy of the system is proposed together with constitutive relations to close the system. In Section~\ref{sec:gradientflow}, the system is showed to be a generalized gradient flow, and in Section~\ref{sec:numerics}, a numerical example compares the newly proposed model with the Cahn-Larch\'e system. 

\section{The derivation of the Cahn-Hilliard-Biot model}\label{sec:model}
We consider a saturated porous medium with one fluid phase, and two solid phases with distinct material properties. The solid phases are modeled by a diffuse interface approach of Cahn-Hilliard type, where surface tension, deformation of the solid material, and pore pressure are acting as driving forces.

Let the medium $\Omega\subset \mathbb{R}^d$ be a bounded domain, $d$ the spatial dimension, and $[0,T]$ be a time interval where $T$ denotes the final time. In the matrix, the smooth phase-field, $\varphi\!:\! \Omega\times[0,T] \rightarrow [-1,1]$, tracks the two phases $\pf=-1$ and $\pf = 1$. We consider linearized elasticity with infinitesimal displacement $\displacement$, and $\|\nabla \displacement\| \ll 1$, the pore pressure is denoted by $p$, and $\bm q$ is the fluid flux.

\subsection{Balance laws}
Balance laws are imposed for each of the three coupled systems. For the phase-field equation, we assume that the phase-change is conserved through a phase-field flux $\bm J$ and reactions $R$,
\begin{equation}
\partial_t\varphi + \nabla \cdot {\bm J} = R.
\end{equation}
The elastic behavior of the material is governed by a quasi-static force balance equation where $\bm \sigma$ denotes the stress tensor and $\bm f$ external body forces \begin{equation}
    -\nabla \cdot \bm \sigma = \bm f.
\end{equation}
Finally, the fluid is assumed to follow a volume balance law with negligible density gradients,
\begin{equation}
    \partial_t \theta + \nabla \cdot \bm q = S_f,
\end{equation}
where $\theta$ is the volumetric fluid content which changes due to the fluid flux $\bm q$ and source $S_f$. 

\subsection{Free energy}
The system is then closed through its free energy together with appropriate constitutive relations. We assume that the energy can be decomposed into three parts; the regularized surface energy, containing chemical energy and interfacial energy between the solid phases, the elastic energy, and the fluid energy
\begin{equation} \label{eq:totenergy}
    \mathcal{E}(\varphi,\bm u, \theta) = \mathcal{E}_\mathrm{ch}(\varphi) + \mathcal{E}_\mathrm{e}(\varphi, \bm u) +\mathcal{E}_\mathrm{f}(\varphi,\bm u, \theta).
\end{equation}

The regularized surface energy \cite{CH1} is given as
\begin{equation}
    \mathcal{E}_\mathrm{ch}(\varphi) := \int_\Omega \Psi(\varphi) + \frac{\gamma}{2}|\nabla\varphi|^2\;dx,
\end{equation}
where deviations from pure phases are penalized through the double-well potential $\Psi(\varphi)$, and transitions between phases are penalized by the second term which is related to the interfacial energy. Here, the parameter $\gamma$ corresponds to interfacial tension between the phases and will account for adhesive and cohesive forces. The double-well potential takes minimal values in the two phases, $\varphi = -1$ and $\varphi = 1$, and is, in this work, given as
\begin{equation}
    \Psi(\varphi) := \frac{1}{4}\left(1-\varphi^2  \right)^2.
\end{equation}

We assume that the elastic energy takes the form that is typical to the Cahn-Larch\'e equations,
\begin{equation}
    \mathcal{E}_\mathrm{e}(\pf,\displacement) = \int_\Omega \frac{1}{2} \big({\bm \varepsilon}(\bm u) - \mathcal{T}(\varphi)\big)\!:\!\mathbb{C}(\varphi)\big({\bm \varepsilon}(\bm u) - \mathcal{T}(\varphi)\big) \;dx,
\end{equation}
where ${\bm \varepsilon}(\bm u) = \frac{1}{2}\left(\nabla \bm u + \nabla\bm u^\top\right)$ is the linearized strain at displacement $\bm u$. The second term, $\mathcal{T}(\varphi)$, is the eigenstrain at $\varphi$ (often called {\it stress-free strain}, or {\it intrinsic strain}) which corresponds to the state of the strain tensor if the material was uniform and unstressed \cite{fratzl}. Moreover, it can be considered to account for swelling effects \cite{areias} and takes different values depending on the solid phase $\pf$. Here, we consider the form $\mathcal{T}(\pf) = \xi \pf\bm I$, where $\xi$ is a swelling parameter. The elastic stiffness tensor $\mathbb{C}(\varphi)$, which can be anisotropic, depends on the phase-field. 

Finally, we consider a natural extension of the classical fluid energy which is given as in \cite{jakubgradientflow} by
\begin{equation}
    \mathcal{E}_\mathrm{f}(\varphi,\bm u, \theta) = \int_\Omega \frac{M(\varphi)}{2}\left(\theta - \alpha(\varphi)\nabla\cdot \bm u\right)^2\; dx
\end{equation}
where both the compressibility parameter  $M(\varphi)$ and the Biot-Willis coupling coefficient $\alpha(\varphi)$ depend on the phase-field $\pf$.

\subsection{Constitutive relations}
Assuming that the phase-field follows Fick's law for non-ideal mixtures, the flux $\bm J$ is proportional to the negative gradient of the chemical potential
\begin{equation}\label{eq:fick}
    \bm J = -m(\pf)\nabla \mu,
\end{equation}
where $m(\pf)$ is the chemical mobility.
The chemical potential $\mu$ is defined to be the variational derivative of the free energy with respect to $\pf$. Here, we denote the variational derivative of $\energy$ with respect to $y$ by $\delta_y \energy$, and standard computations yield \begin{equation}\label{eq:mu}
    \mu := \delta_\pf \energy = \Psi'(\varphi) - \Delta \varphi + \delta_\varphi\mathcal{E}_\mathrm{e}(\varphi, \bm u) + \delta_\varphi \mathcal{E}_\mathrm{f}(\varphi, \bm u, \theta),
\end{equation}
where zero Neumann or periodic boundary conditions have been applied to $\pf$,
\begin{equation}
    \delta_\varphi\mathcal{E}_\mathrm{e}(\varphi, \bm u) = \frac{1}{2}\left({\bm \varepsilon}(\bm u) - \mathcal{T}(\varphi)\right)\!:\!\mathbb{C}'(\varphi)\left({\bm \varepsilon}(\bm u) - \mathcal{T}(\varphi)\right) - \mathcal{T}'(\varphi)\!:\!\mathbb{C}(\pf)\left({\bm \varepsilon}(\bm u) - \mathcal{T}(\varphi)\right),
\end{equation}
and
\begin{equation}\label{eq:derenergyfluid}
    \delta_\varphi \mathcal{E}_\mathrm{f}(\varphi, \bm u, \theta) = \frac{M'(\varphi)}{2}(\theta-\alpha(\varphi)\nabla\cdot \bm u)^2 - M(\pf)(\theta-\alpha(\pf)\nabla\cdot \displacement)\alpha'(\varphi)\nabla\cdot\bm u.
\end{equation}
According to thermodynamical principles \cite{coussy}, we define the stress tensor to be the rate of change of energy with respect to strain 
\begin{equation}\label{eq:stress}
    \bm \sigma := \delta_{\bm\varepsilon}\mathcal{E} = \mathbb{C}(\varphi)\left({\bm \varepsilon}(\bm u)-\mathcal{T}(\varphi)\right) - M(\varphi)\alpha(\varphi)\left(\theta-\alpha\nabla\cdot \bm u\right)\bm I,
\end{equation}
and the pore pressure $p$ to be the rate of change of energy with respect to volumetric fluid content 
\begin{equation}\label{eq:pressure}
p := \delta_\theta \energy = M(\pf)\left(\theta-\alpha(\varphi)\nabla\cdot\displacement\right).
\end{equation}
Finally, the flow through the porous medium is assumed to follow Darcy's law
\begin{equation}\label{eq:darcy}
    \bm q = -\kappa(\varphi)\nabla p,
\end{equation}
where the permeability $\kappa(\pf)$ is assumed to depend on the solid phase.

Combining the balance laws with the constitutive relations, and making the identification \eqref{eq:pressure} in \eqref{eq:derenergyfluid} and \eqref{eq:stress}, the Cahn-Hilliard-Biot model becomes
\begin{eqnarray}
\partial_t \varphi - \nabla \cdot (m(\pf) \nabla \mu) &=& R \label{eq:ch1}\\
\mu +\gamma\Delta \varphi - \Psi'(\varphi) - \delta_\varphi\mathcal{E}_\mathrm{e}(\varphi, \bm u) - \delta_\varphi \mathcal{E}_\mathrm{f}(\varphi, \bm u, p) &=& 0\label{eq:ch2}\\
-\nabla \cdot\left(\mathbb{C}(\varphi)\left({\bm \varepsilon}(\bm u)-\mathcal{T}(\varphi)\right)\right) + \nabla \left(\alpha(\varphi)p \right) &=& \bm f\label{eq:elasticity}\\\label{eq:flow}
\partial_t\left(\frac{p}{M(\varphi)} + \alpha(\varphi)\nabla\cdot \bm u\right) + \nabla\cdot \bm q &=& S_f\\
\bm q +
\kappa(\varphi)\nabla p&=& 0, \label{eq:darcyflow}
\end{eqnarray}
equipped with suitable boundary and initial conditions. 

\section{The Cahn-Hilliard-Biot model as a generalized gradient flow}\label{sec:gradientflow}
In this section, we identify the proposed Cahn-Hilliard-Biot model \eqref{eq:ch1}--\eqref{eq:darcyflow} as a generalized gradient flow, which in contrast to regular gradient flows allow for non-quadratic and even degenerate dissipation potentials. By making this identification for the newly proposed model, a wide toolbox of well-posedness analysis \cite{colli1992,jakubgradientflow}, numerical error analysis \cite{savare2000, bartels2014} and numerical solution algorithms \cite{jakubgradientflow, jungel2019} are made available, which will be a valuable asset for further study. A generalized gradient flow takes the form 
\begin{equation}\label{eq:gradientflow}
    \D_{\partial_t\bm z} \mathcal{R}(\partial_t\bm z, \bm z)  = -\mathcal{D}_{\bm z} \mathcal{E}(\bm z) + \mathcal{P}_{\mathrm{ext}},
\end{equation}
where $\bm z$ is a state variable, $\mathcal{R}$ is a dissipation potential, $\mathcal{E}$ is the energy at state $\bm z$, $\D_{\bm x}$ is the Gateaux gradient with respect to $\bm x$, and $\mathcal{P}_{\mathrm{ext}}$ corresponds to external forces. Alternatively, one can reformulate the generalized gradient flow and split between states evolving with ($\bm z_\mathrm{d}$) and without ($\bm z_\mathrm{df}$) dissipation to get the constrained minimization problem
\begin{eqnarray}
    \bm z_\mathrm{df} &=& \argmin_{\bm s_{\mathrm{df}}}\left\{\energy(\bm s_\mathrm{df})-\langle\mathcal{P}_{\mathrm{ext},\mathrm{df}},\bm s_\mathrm{df}\rangle \right\} 
\\
    (\partial_t{\bm z_\mathrm{d}}, \mathcal{F}) &=& \argmin_{\bm s_\mathrm{d}, \bm l}\Big\{\tilde{\R}(\bm l, \bm z_\mathrm{d}) + \langle \D_{\bm z_\mathrm{d}}\mathcal{E}(\bm z_\mathrm{d}),\bm s_\mathrm{d}\rangle - \langle\mathcal{P}_{\mathrm{ext},\mathrm{d}},\bm s_\mathrm{d}\rangle \Big\}
\end{eqnarray}
subject to $\bm s_\mathrm{d} + \nabla \cdot \bm l = \bm S$, where $\mathcal{R}(\partial_t{\bm z_\mathrm{d}},\bm z_{\mathrm{d}}) = \tilde{\mathcal{R}}(\mathcal{F}, \bm z_{\mathrm{d}})$, $\langle\cdot,\cdot\rangle$ is the canonical inner-product, and the balance law $\partial_t{\bm z_d} + \nabla \cdot \mathcal{F} = \bm S$ with flux $\mathcal{F}$, and source $\bm S$ holds.


For the Cahn-Hilliard-Biot system, consider the state variables $\bm z = (\varphi, \bm u, \theta)$, the energy $\energy(\bm z)$  from \eqref{eq:totenergy}, and the state-dependent dissipation potential 
\begin{equation}
    \R(\bm J, \partial_t{\bm u}, \bm q, \pf) := {\R}_\mathrm{ch}(\bm J,\pf) + \mathcal{R}_\mathrm{e}(\partial_t{\bm u}) + {\mathcal{R}}_\mathrm{f}(\bm q,\pf),
\end{equation}
with $$\mathcal{R}_\mathrm{ch}(\bm J,\pf) := \int_\Omega \frac{1}{2m(\pf)}|\bm J|^2  \; dx, \quad
\mathcal{R}_\mathrm{e}(\partial_t{\bm u}) := 0,\quad \mathrm{and}\quad
\mathcal{R}_\mathrm{f}(\bm q,\pf) := \int_\Omega \frac{1}{2\kappa(\pf)}|\bm q|^2\; dx$$
together with the conservation laws 
\begin{equation}
\partial_t{\pf} +\nabla\cdot \bm J =R \label{eq:balance}\quad \mathrm{and} \quad\partial_t{\theta} +\nabla\cdot \bm q = S_f.
\end{equation}
As the deformation is assumed to be dissipation free, the generalized gradient flow reads: Find $\pf$, $\bm u$, and $\theta$ such that
\begin{eqnarray}
\displacement\hspace{-0.2cm} &=& \hspace{-0.2cm}\argmin_{\bm w} \Big\{\mathcal{E}(\pf,\bm w, \theta)-\langle\mathcal{P}_{\mathrm{ext},\mathrm{e}},\bm w\rangle\Big\}  
\\
\hspace{-0.3cm}(\partial_t{\pf}, \partial_t{\theta}, \bm J, \bm q) \hspace{-0.2cm}&=&\hspace{-0.2cm} \argmin_{\eta, s,\bm l, \bm v} \Big\{{\mathcal{R}}_{\mathrm{ch}}(\bm l, \pf) +\langle\D_\pf\mathcal{E}(\pf, \bm u,\theta),\eta\rangle +{\mathcal{R}}_{\mathrm{f}}(\bm v, \pf) +\langle\D_\theta\mathcal{E}(\pf, \bm u,\theta),s\rangle + \langle\mathcal{P}_{\mathrm{ext},\mathrm{f}},s\rangle
\Big\}
\end{eqnarray}
subject to $\eta+\nabla\cdot \bm l = R$ and $s+\nabla \cdot \bm v = S_f$ with balance laws \eqref{eq:balance}, $\langle\mathcal{P}_{\mathrm{ext},\mathrm{e}},\bm w\rangle := \int_\Omega \bm f\cdot \bm w \; 
dx$ and $\mathcal{P}_{\mathrm{ext},\mathrm{f}}$ corresponding to external forces related to the fluid (e.g., boundary conditions or gravitational force).
Calculating optimality conditions, and substituting the phase-field flux $\bm J$ by the chemical potential $\mu$ through Fick's law \eqref{eq:fick}, and the volumetric fluid content $\theta$ with the fluid pressure $p$ through the relation \eqref{eq:pressure}, one obtains the variational form of the system \eqref{eq:ch1}--\eqref{eq:darcyflow}.

\section{Numerical example}\label{sec:numerics}
Here, we present a numerical example that emphasizes the need for the Cahn-Hilliard-Biot model. We compare a simulation of the Cahn-Hilliard-Biot model with and without a pressure boundary condition acting as an external force (in order to enforce flow in the domain), with a Cahn-Larch\'e simulation (Cahn-Hilliard coupled with only elasticity). The example clearly shows that when the fluid flow is dominant, it also plays a crucial role in the evolution of the phase-field. However, in regimes with little, to no flow, the phase-field is unaffected compared to the Cahn-Larch\'e model. 

\begin{figure}
\centering
\begin{minipage}{0.5\textwidth}
    \begin{subfigure}{0.24\textwidth}
        \includegraphics[width = \textwidth]{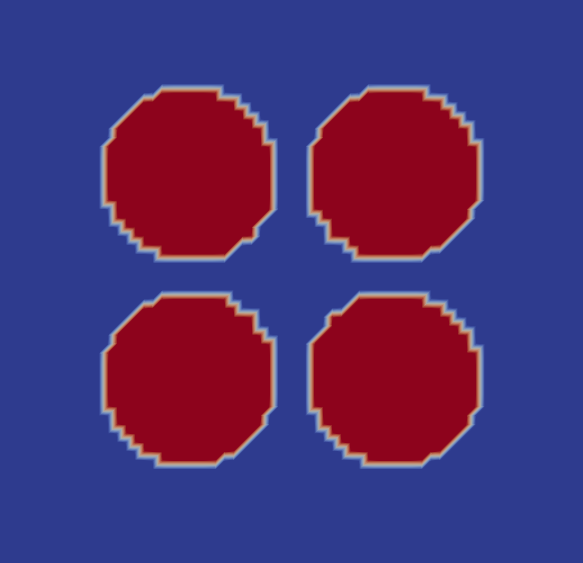}
        \caption{$t = 0$}
        \label{fig:CHBPDt0}
    \end{subfigure}
    \hfill
    \begin{subfigure}{0.24\textwidth}
        \includegraphics[width = \textwidth]{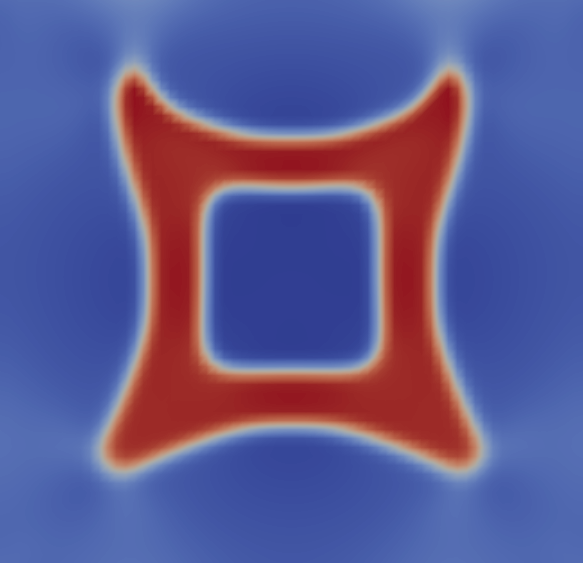}
        \caption{$t = 0.075$}
        \label{fig:CHBPDt75}
    \end{subfigure}
    \hfill
    \begin{subfigure}{0.24\textwidth}
        \includegraphics[width = \textwidth]{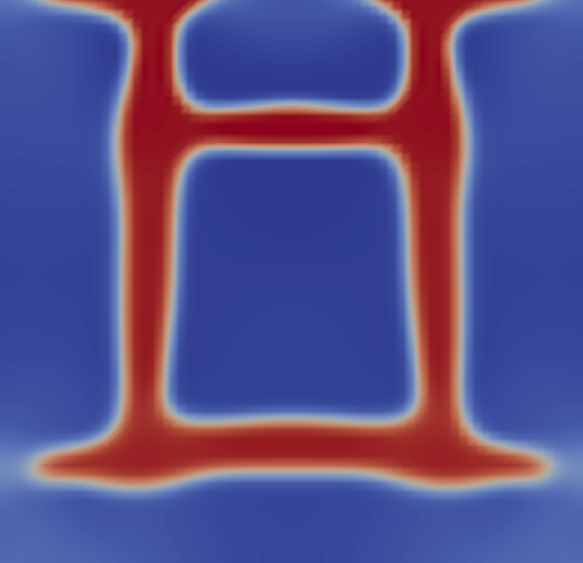}
        \caption{$t = 0.50$}
        \label{fig:CHBPDt500}
    \end{subfigure}
    \hfill
    \begin{subfigure}{0.24\textwidth}
        \includegraphics[width = \textwidth]{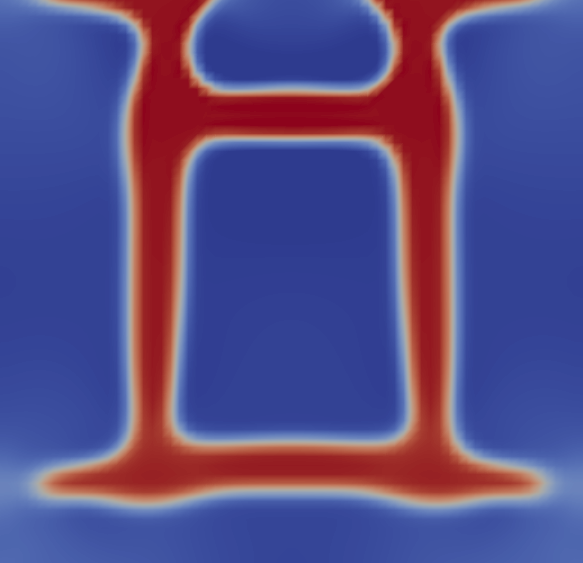}
        \caption{$t = 1.0$}
        \label{fig:CHBPDt1000}
    \end{subfigure}
    \hfill
    \begin{subfigure}{0.24\textwidth}
        \includegraphics[width = \textwidth]{figs-circle/CHE0.png}
        \caption{$t = 0$}
        \label{fig:CHBt0}
    \end{subfigure}
    \hfill
    \begin{subfigure}{0.24\textwidth}
        \includegraphics[width = \textwidth]{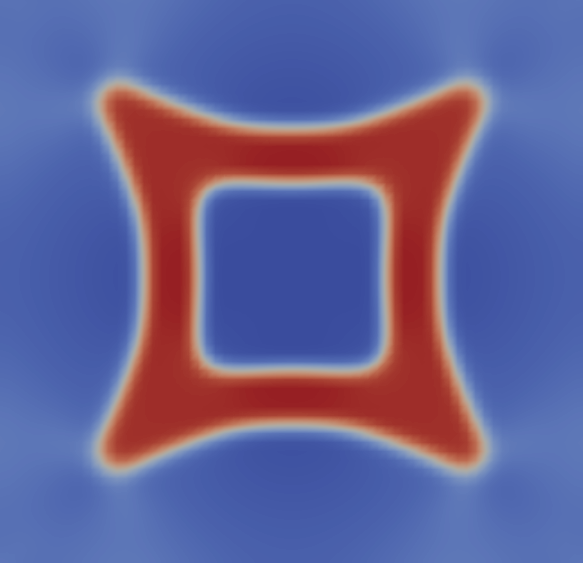}
        \caption{$t = 0.075$}
        \label{fig:CHBt75}
    \end{subfigure}
    \hfill
    \begin{subfigure}{0.24\textwidth}
        \includegraphics[width = \textwidth]{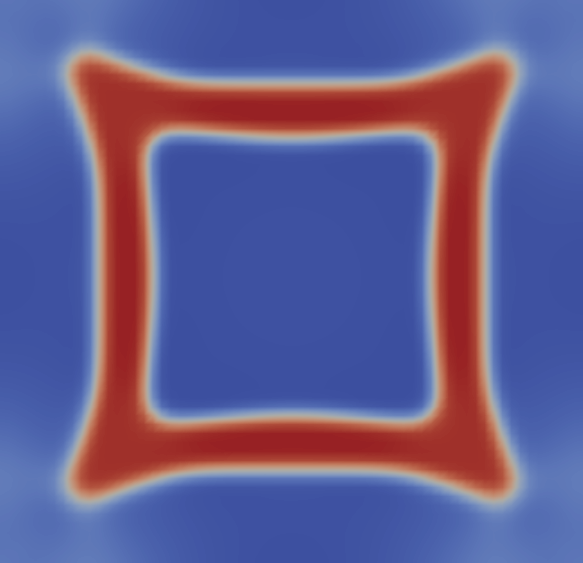}
        \caption{$t = 0.50$}
        \label{fig:CHBt500}
    \end{subfigure}
    \hfill
    \begin{subfigure}{0.24\textwidth}
        \includegraphics[width = \textwidth]{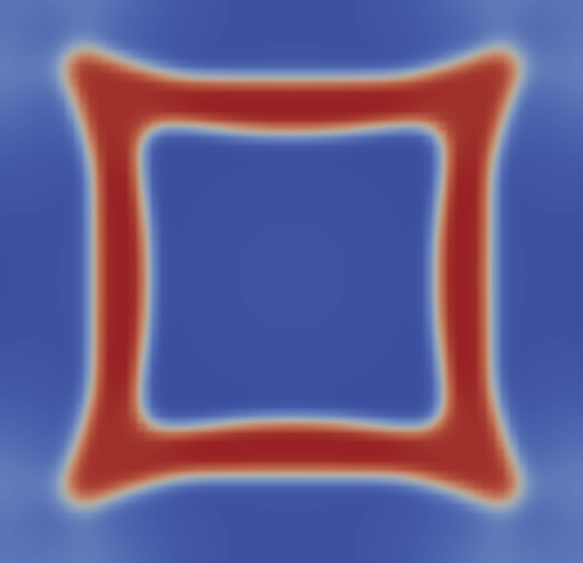}
        \caption{$t = 1.0$}
        \label{fig:CHBt1000}
    \end{subfigure}
        \begin{subfigure}{0.24\textwidth}
        \includegraphics[width = \textwidth]{figs-circle/CHE0.png}
        \caption{$t = 0$}
        \label{fig:CHEt0}
    \end{subfigure}
    \hfill
    \begin{subfigure}{0.24\textwidth}
        \includegraphics[width = \textwidth]{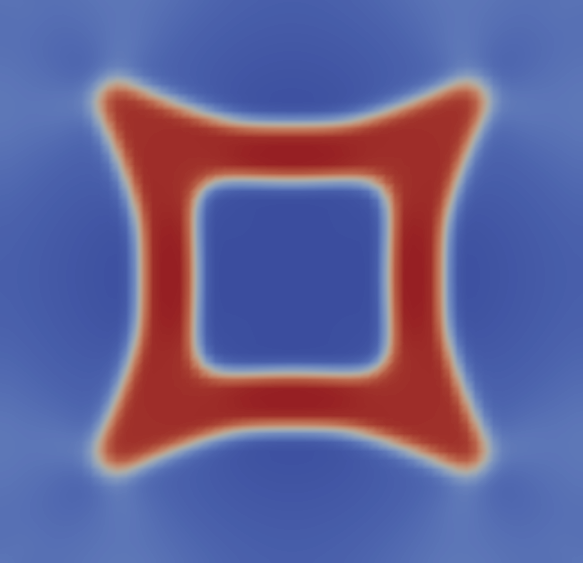}
        \caption{$t = 0.075$}
        \label{fig:CHEt75}
    \end{subfigure}
    \hfill
    \begin{subfigure}{0.24\textwidth}
        \includegraphics[width = \textwidth]{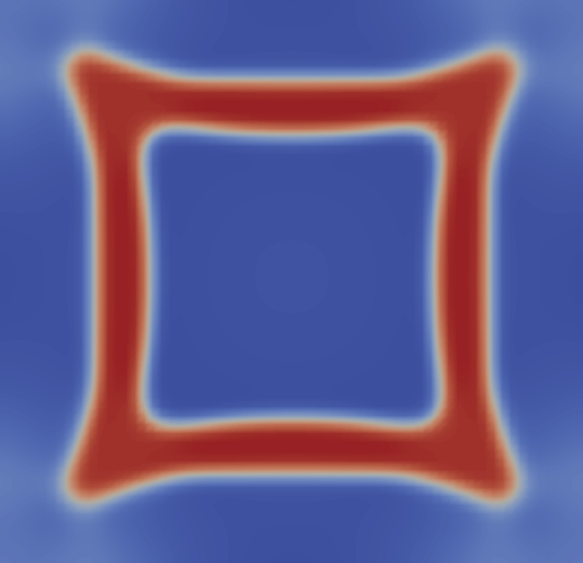}
        \caption{$t = 0.50$}
        \label{fig:CHEt500}
    \end{subfigure}
    \hfill
    \begin{subfigure}{0.24\textwidth}
        \includegraphics[width = \textwidth]{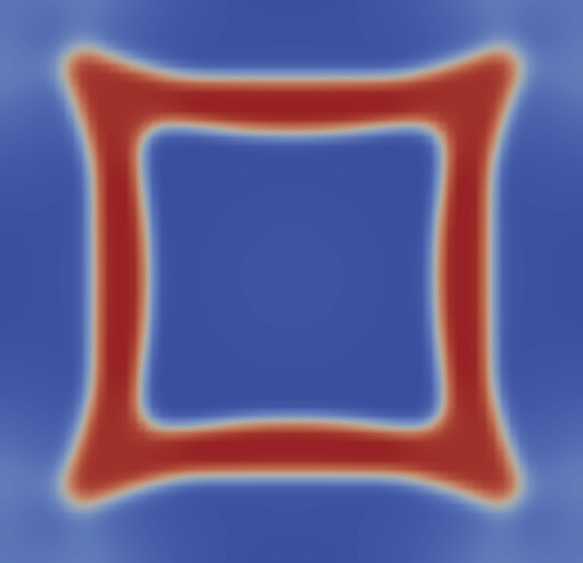}
        \caption{$t = 1.0$}
        \label{fig:CHEt1000}
    \end{subfigure}
    \label{fig:CHB}
        \hspace{-0.47cm}\includegraphics[width=1.11\textwidth]{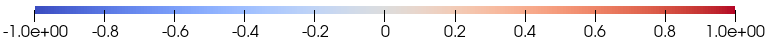}
    \end{minipage}
    \begin{minipage}{0.45\textwidth}
    \begin{subfigure}{\textwidth}


    \caption{Energies for all three simulations.}
    \label{fig:energy}
    \end{subfigure}
    \end{minipage}
    \caption{(a) -- (l): the solution at time $t$ for the phase-field $\pf$. (a) -- (d): Cahn-Hilliard-Biot with $p=0.25$ on the top, (e) -- (h): Cahn-Hilliard-Biot with zero pressure BC, (i) -- (l): Cahn-Larch\'e. (m): system energy (with external contributions). PD is Cahn-Hilliard-Biot with $p=0.25$ on the top, CHB is Cahn-Hilliard-Biot with zero pressure BC and CHE is Cahn-Larch\'e.}
\end{figure}

We consider a unit square domain where four circular shapes of phase $\pf = 1$ are surrounded by phase $\pf = -1$ initially, see Figure~\ref{fig:CHBPDt0},\ref{fig:CHBt0},\ref{fig:CHEt0}. For both pressure and displacement, we apply zero initial data. The variational system \eqref{eq:ch1}--\eqref{eq:darcyflow} is discretized in time by a semi-implicit Euler method, where the deviation from fully implicit Euler is an application of the first order convex splitting method of the double-well potential $\Psi(\pf)$ as proposed in \cite{eyre1998}. The three-way coupled nonlinear system is then solved by an iterative decoupling scheme, starting with the Cahn-Hilliard subsystem \eqref{eq:ch1}--\eqref{eq:ch2}, then elasticity \eqref{eq:elasticity}, and finally,  flow \eqref{eq:flow}--\eqref{eq:darcyflow}. 
The Cahn-Hilliard subsystem \eqref{eq:ch1}--\eqref{eq:ch2} is discretized in space with bilinear rectangular finite elements for both phase-field $\pf$ and chemical potential $\mu$, and the nonlinear equations are solved by a Newton method in each iterative  decoupling-iteration.
The flow subsystem \eqref{eq:flow}--\eqref{eq:darcyflow} is discretized in space by lowest-order Raviart-Thomas elements, RT0, for the flux and constant elements for pressures, and the elasticity equation \eqref{eq:elasticity} is discretized with bilinear finite elements. We have used modules from the DUNE project, specifically dune-functions \cite{dune-functions}, for the implementation.

The material parameters can be found in Table~\ref{tab:1}, and the permeability $\kappa(\pf)$, compressibility $M(\pf)$, Biot-Willlis coefficient $\alpha(\pf)$ and elasticity tensor $\mathbb{C}(\pf)$ are depending on the phase-field through the interpolation function $\pi(\pf)$; $\kappa(\pf) = \kappa_{-1} + \pi(\pf)(\kappa_1-\kappa_{-1})$, $M(\pf) = M_{-1} + \pi(\pf)(M_1-M_{-1})$, $\alpha(\pf) = \alpha_{-1} + \pi(\pf)(\alpha_1-\alpha_{-1})$ and $\mathbb{C}(\pf) = \mathbb{C}_{-1} + \pi(\pf)(\mathbb{C}_1 -\mathbb{C}_{-1})$. Here, we choose\\
\begin{minipage}{0.45\textwidth}
\begin{equation*}
    \pi(\pf) = 
    \begin{cases} 
    0,\quad &\pf<-1\\
    \frac{1}{4}\left(-\pf^3 + 3\pf +2 \right), \quad &\pf\in [-1,1]\\
    1,\quad &\pf>1
    \end{cases},
\end{equation*}
\end{minipage}
\begin{minipage}{0.45\textwidth}
\begin{equation*}
    \mathbb{C}_{-1} = \begin{pmatrix} 
    4 & 2 & 0 \\
    2 & 4 & 0 \\
    0 & 0 & 8\end{pmatrix}, \qquad 
    \mathbb{C}_{1} = \begin{pmatrix} 
    1 & 0.5 & 0 \\
    0.5 & 1 & 0 \\
    0 & 0 & 2\end{pmatrix},
\end{equation*}
\end{minipage}
\\
as in \cite{garcke2005CHENumerics}, with the two elasticity tensors written in Voigt notation in two spatial dimensions.
Zero Neumann boundary conditions are applied to both the phase-field and the chemical potential, while the displacement is equipped with zero Dirichlet conditions on the entire boundary. For the flow subsystem, we enforce a pressure drop from $p=0.25$ to $p=0$ from top to bottom while no-flow conditions are applied on the left and right parts of the boundary.

\begin{table}[ht]
\centering
\begin{tabular}{c|c|c||c|c|c}
Parameter name & Symbol & Value & Parameter name & Symbol & Value\\
\hline
Chemical mobility & $m$ & 1 & Biot-Willis parameters &  $\alpha_{-1}$, $\alpha_1$ & 1, 0.5 \\
Interfacial tension & $\gamma$ & 1e-4 & Permeabilities & $\kappa_{-1}$, $\kappa_{1}$& 0.1, 1 \\
Swelling parameter & $\xi$ & 0.3 & Mesh size diameter & $h$ & $\frac{\sqrt{2}}{65}$ \\
Compressibilities &  $M_{-1}, M_1$ & 1, 0.1 & Time step size & $\tau$ & 1e-3\\
\end{tabular}
\caption{Table of simulation parameters.}
\label{tab:1}
\end{table}

In Figure~\ref{fig:CHBPDt0}--\ref{fig:CHBPDt1000}, the phase-field function $\pf$ is plotted after a series of time steps for the Cahn-Hilliard-Biot model with a drop in pressure from $p=0.25$ to $p=0$ from top to bottom. In Figure~\ref{fig:CHBt0}--\ref{fig:CHBt1000} the solution is plotted at the same time steps, but with zero pressure on the entire boundary, and similarly in Figure~\ref{fig:CHEt0}--\ref{fig:CHEt1000} the plots are from a simulation of the Cahn-Larch\'e system. We observe that when the flow is prominent in the simulation the phase-field is also significantly affected and takes a directional preference to that of the flow direction. When, on the other hand, the system merely is filled with a fluid that has no driving force in itself, the phase-field evolution is close to unaffected compared to the system without a fluid. We emphasize also that the system energies (including external forces) are decreasing over the scope of the simulation, as is expected from dissipative systems of gradient flow type. This is showed in Figure~\ref{fig:energy}, where the energy is a combination of the free energy of the system \eqref{eq:totenergy}, and the external forces applied through the pressure boundary condition, $\mathcal{E}_\mathrm{Tot} = \mathcal{E}(\varphi, \bm u, p) - \int_{\Gamma_\mathrm{Top}} p_{\mathrm{Top}} (\bm q\cdot \bm n) \;dx$, $\bm n$ being the outwards pointing normal vector.

\section{Conclusions}
The Cahn-Hilliard-Biot system was derived through balance laws and constitutive relations, i.e., Fick's law for the phase-field, and Darcy's law for the fluid flow. Key quantities are defined, following thermodynamical principles, as rates of change of the free energy. The equations feature a three-way coupling, and the impact from flow to the phase-field was showed to be significant through a numerical example; the phase-field does not only evolve as it would through the Cahn-Larch\'e equations, but its evolution is aligned and magnified in the flow direction. Moreover, we showed that the system follows a generalized gradient flow framework and that the energy dissipates numerically as expected. By this, we lay the groundwork for a general model, showing numerical properties and highlighting important coupling terms, that can be further tailored and studied depending on the specific application in mind.
%

\bibliographystyle{unsrt}
{\bibliography{chb-letter-arxiv.bib}}

\end{document}